\documentstyle[hep93,twocolumn]{article}
\include{figures}
\begin{document}
\twocolumn[


\begin{center}
{\Large \bf Hadron Production Ratios as Probes\\
\vspace{.3cm}
of Confinement and Freeze-Out}
\\
\vspace{.3cm}
{\bf E. Suhonen}$^{a}$, {\bf J. Cleymans}$^{b}$,
{\bf K. Redlich}$^{c}$, {\bf H. Satz}$^{d}$\\
\vspace{.5cm}
$^{a}$Department of Theoretical Physics,
University of Oulu, FIN-90570 Oulu, Finland.\\
$^{b}$Department of Physics, University of Cape Town,
Rondebosch 7700, South Africa.\\
$^{c}$Fakult\"{a}t f\"{u}r Physik, Universit\"{a}t
Bielefeld, D-33501 Bielefeld, Germany.\\
$^{d}$Theory Division, CERN, CH-1211 Geneva 23,
Switzerland.
\end{center}
\vspace{.3cm}
\begin{center}
\noindent
Particle production in central S-A collisions at 200 GeV/A energy is
analysed within a thermal model. Present data imply that the strange
particles freeze out at a higher temperature than the non-strange
particles and that the strangeness saturation is incomplete.
\vspace{.3cm}
\end{center}
]

\underline{Introduction}: $\;$
A significant enhancement of strange particle production has been
observed in high energy heavy ion collisions in comparison to proton-
proton
collisions. This indicates a higher level of equilibration in
nuclear collisions where volume, life-time and energy density are
increased. Strongly interacting matter produced initially expands
rapidly, cools down and breaks up into observable hadrons. It is not
excluded that a quark-gluon plasma phase in which the colour degrees
of freedom are deconfined is achieved at an early stage of evolution.
Soft processes involved in the transition from quarks and gluons to
hadrons are, if at all, only poorly understood. It is possible that
the system expands too fast to retain equilibrium and a quark-gluon
plasma
suddenly disintegrates into the final hadrons. In this case the
observed
hadrons could carry information on parameters and properties of quark-
gluon
plasma and on confinement \cite{ref1,ref2}. We assume here a smooth
evolution in which a hadron gas thermalises before the final hadrons
stop interacting. Regardless of the exact nature of the produced
matter,
the observed hadrons will reflect the properties of the
last thermal state before the freeze-out, the equilibrium hadron gas.
In the present paper we shall study data on hadron production ratios
within the framework of a thermal model to determine the statistical
properties of their sources, i.e. of particle freeze-out states.


\newpage
\normalsize
\underline{Thermal Model}: $\;$
The state of an equilibrium hadron gas is specified by three
parameters, the temperature $T$, the baryon chemical potential
$\mu_{B}$ and the strangeness chemical potential $\mu_{s}$. The
requirement of vanishing overall strangeness fixes one of these
parameters, e.g. $\mu_{s}$, and the two remaining parameters,
$T$ and $\mu_{B}$, fully determine the local state in the thermal
model. For simplicity, we present the formulae in Boltzmann
statistics although in the actual calculations we have worked with
the quantum statistics. The partition function of the hadron gas is
given by \cite{ref1}
\begin{equation}
\makebox{ln} \; Z(T,\mu_{B},\mu_{s}) = \sum_{i}
\left[ W_{i}^{m} + \left( \lambda_{B}^{B_{i}} \lambda_{S}^{-S_{i}}
+ \lambda_{B}^{-B_{i}} \lambda_{S}^{S_{i}} \right) W_{i} \right] .
\end{equation}
Here the first term refers to non-strange mesons and the second term
to particles which carry baryon numbers $B_{i}$ and strangeness
$S_{i}$. The fugacities related to the baryon number and strangeness
are $\lambda_{B} \equiv \makebox{exp} (\mu_{B}/T)$ and
$\lambda_{S} \equiv \makebox{exp} (\mu_{s}/T)$. The phase space
factor $W_{i}$ is of the form
\begin{equation}
W_{i} = \frac{d_{i} VT m_{i}^{2}}{2 \pi^{2}} \; K_{2} \;
        \left( \frac {m_{i}}{T} \right)
\end{equation}
where $d_{i}$ denotes the degeneracy and $m_{i}$ the mass of the
hadron state $i$, $V$ is the volume of the system and $K_{2}$ the
modified Bessel function of the second type. We have included in
eq.(1) all well established resonances up to mass of 2 GeV. The
thermal contribution of the particle multiplicity $N_{i}^{th} =
W_{i}$ calculated from eq.(1) has to be added by the resonance
contributions to get the particle multiplicity,
\begin{equation}
N_{i} = W_{i} + \sum_{j} \; \Gamma_{ij} \; W_{j} .
\end{equation}
Here $\Gamma_{ij}$ is the branching ratio of the decay of resonance
$j$ to particle $i$. If both thermal and chemical equilibrium were
established
and if there were a unique freeze-out for different hadron species,
then all hadron production ratios would be determined using the
values of $T$ and $\mu_{B}$ fixed by two measured ratios.

\underline{Data Interpretation}: $\;$
Data we consider here are from sulphur collisions at 200A
GeV energy with tungsten- \cite{ref3,ref4}, silver- \cite{ref5} and
lead- \cite{ref2,ref6,ref7} targets. They are measured in the backward
hemisphere, 2.3 $<$ y $<$ 3.0, where the production ratios do not
depend much
on the target size. The WA85 experiment \cite{ref3,ref4} provides us
with
the following strange baryon and antibaryon ratios:
$\bar{\Lambda}/\Lambda = 0.2
\pm 0.01, \overline{\Xi^{-}}/\Xi^{-} = 0.45 \pm 0.05$, $\Xi^{-
}/\Lambda
= 0.095 \pm 0.006$, $\overline{\Xi^{-}}/\bar{\Lambda} = 0.21 \pm 0.02$
and $\overline{\Omega^{-}}/\Omega^{-} = 0.57 \pm 0.41$. The ratios for
$\Xi^{-}/\Lambda$ and $\overline{\Xi^{-}}/\bar{\Lambda}$ have been
corrected for the $p_{T}$ cut while the result for
$\overline{\Omega^{-}}/\Omega^{-}$
is preliminary and uncorrected for the acceptance. The first two
ratios
lead to the narrow bands in the $T-\mu_{B}$ plane which cross each
other in the small region of $T \simeq$ (190 $\pm$ 15) MeV and
$\mu_{B} \simeq$ (240 $\pm$ 40) MeV, as shown in Figure 1. The
crossing region corresponding to the ratios for $\Xi^{-}/\Lambda$
and $\overline{\Xi^{-}}/\bar{\Lambda}$ is quite different, as also
shown in
Figure 1. The thermal model in its original form thus turns out to be
\begin{figure}[h]
\vspace{8.5cm}
\end{figure}
too idealized to explain WA85 data. A remedy we use is to leave out
the assumption from the complete chemical equilibrium between strange
and non-strange hadrons. In fact it has been suggested, on the basis
of small cross-sections of strange particle production, that the
strange particle phase space can reach only partial saturation
\cite{ref2,ref8,ref9}. The exchange processes among the strange
particle species are faster than strangeness producing processes and
therefore the strange particles are assumed to be in equilibrium
relative to each other but in relation to the non-strange particles
they may be suppressed by a phase space saturation factor $\gamma_{s}
< 1$
\cite{ref8}. This is achieved in our model by multiplying both
$\lambda_{S}$ and $\lambda_{S}^{-1}$ in eq.(1) by a parameter
$\gamma_{S}$
which then yields for the multiplicity of particle species $i$,
\begin{equation}
N_{i} = \gamma_{S}^{S_{i}} W_{i} + \sum_{j} \; \gamma_{S}^{S_{j}} \;
\Gamma_{ij} \; W_{j} .
\end{equation}
The ratios for $\bar{\Lambda}/\Lambda$ and $\overline{\Xi^{-}}/\Xi^{-
}$
are not changed by the modification while the results for
$\Xi^{-}/\Lambda$ and $\overline{\Xi^{-}}$ become multiplied by
factor
$\gamma_{S}$. As shown in the figure, the thermal model with
$\gamma_{S} = 0.7$ is compatible with the WA85 measurements on
$\Lambda$, $\bar{\Lambda}$, $\Xi^{-}$ and $\overline{\Xi^{-}}$
production. We notice, however, that the three independent data
points have been used to fix three statistical parameters
$T$, $\mu_{B}$ and $\gamma$ only. More data is needed to justify
the validity of the model. The model prediction for
$\overline{\Omega^{-}}/\Omega^{-}$ (1 $\pm$ 0.3) is higher than the
preliminary experimental ratio. On the other hand, the results for
$K_{s}^{0}/\Lambda$ and $K^{+}/K^{-}$, $K_{s}^{0}/\Lambda \cong$
1.2 $\pm$ 0.5 (2.3 $<$ y $<$ 2.8) and $K^{+}/K^{-} \simeq$ 1.5 $\pm$
0.5
(y $\simeq$ 2.3) obtained from NA35 measurements on rapidity
distributions in S-Ag collisions \cite{ref5} are well predicted by
the model with the above values of $T$ and $\mu_{B}$.
\begin{figure}[h]
\vspace{8.5cm}
\end{figure}
Having extracted the freeze-out parameters for strange particles to
be $T \simeq$ 190 MeV, $\mu_{B} \simeq$ 240 MeV and
$\gamma_{S} \simeq$ 0.7 we turn our attention to the non-strange
hadron production. The non-strange particles dominate the charged
particle multiplicities which have been measured by EMU05 \cite{ref2}
and NA35 \cite{ref5} collaborations. From EMU05 we have a result for
the charge asymmetry ratio, $D_{Q} = (h^{+} - h^{-})/(h^{+} + h^{-})$
= 0.088 $\pm$ 0.007, measured in S-P collisions using the same
rapidity window as for WA85 results. The ratio $D_{Q}$ is closely
related to the entropy per baryon (S/B), $D_{Q}$(S/B) $\simeq$ 4.5,
from which one obtains S/B $\simeq$ 50 \cite{ref2}. In order to see
whether this value is consistent with the thermal model we show in
Fig.3 the entropy per baryon in the hadron gas for several different
\begin{figure}[h]
\vspace{8.5cm}
\end{figure}
temperatures. It is seen that at $T$ and $\mu_{B}$ required by
strange particle ratios, S/B $\simeq$ 30, which is lower than the
measured value 50. The measured value would correspond to temperature
110 MeV $< T <$ 140 MeV as seen in Fig.3. Almost the same
temperature and chemical potential, 120 MeV $< T <$ 140 MeV and
200 MeV $< \mu_{B} <$ 270 MeV, can also be extracted from NA35
measurements on $h^{-}/(p - \bar{p})$ in S-Ag collisions at 2.3
$< y <$ 2.8 \cite{ref5}. Our interpretation for these results is that
there is no unique freeze-out for strange and non-strange particles.
The sequential freeze-out has immediate other experimental
consequences. It implies a different freeze-out radius for kaons
$R_{K}$ than for pions $R_{\pi}$. The interferometry studies of NA44
on S-Pb collisions indeed measure $R_{K} < R_{\pi}$ \cite{ref6,ref7}.
{}From free mean path arguments \cite{ref1}, assuming $\pi\pi$-cross-
section to be twice as big as $\pi K$-cross-section, one gets
$R_{K}/R_{\pi} \simeq$ 0.7 which is rather well in agreement with
NA44
results \cite{ref6,ref7}. For an isentropic expansion the freeze-out
radius is inversely proportional to the freeze-out temperature; hence
$T_{K} \simeq$ 190 MeV would imply $T_{\pi} \simeq$ 130 MeV. This is
consistent with the results of our analysis.

\underline{Conclusions}: $\;$
The production rates of different hadrons provide tools for the study
of hadronisation and freeze-out stages in high energy heavy ion
collisions. The measured ratios of strange and non-strange hadrons
allowed us to determine the freeze-out parameters. Analysing data of
several CERN collaborations (WA85, NA35, EMU05 and NA44) within a
thermal model indicated that the strange particles freeze-out at
higher temperature ($T \cong$ 190 MeV) and at the same chemical
potential ($\mu_{B} \simeq$ 240 MeV) than the non-strange particles
($T \simeq$ 130 MeV). The sequential freeze-out is consistent with
the difference in the mean free paths of kaons and pions in the
medium. The saturation of the strangeness was found to be
incomplete.\\

\end{document}